\documentclass{elsart41}
\usepackage{graphicx}
\usepackage{amssymb}
\usepackage{times}
\begin{document}

\begin{frontmatter}
\title{
  Conductance of a double quantum dot with correlation--induced 
  wave function renormalization
}
%

\author{Adam Rycerz\corauthref{Rycerz}},
\ead{adamr@th.if.uj.edu.pl}
\author{Jozef Spa{\l}ek}
\address{
  Marian Smoluchowski Institute of Physics, Jagiellonian University, 
  Reymonta 4, 30--059 Krak\'{o}w, Poland
}

\corauth[Rycerz]{
  Corresponding author. Present address: \textit{Instituut--Lorentz, 
    Universiteit Leiden, P.O.\ Box 9506, NL--2300 RA Leiden, The Netherlands}
}

\begin{abstract}
The zero--temperature conductance of diatomic molecule, modelled 
as a correlated double quantum dot attached to noninteracting leads
is investigated. We utilize the Rejec--Ram\v{s}ak formulas, relating 
the linear--response conductance to the ground--state energy dependence
on magnetic flux within the framework of EDABI method, which combines exact 
diagonalization with \emph{ab initio} calculations. The single--particle
basis renormalization leads to a strong particle--hole asymmetry, of the
conductance spectrum, absent in a standard parametrized model study. 
We also show, that the coupling to leads $V\approx 0.5t$ ($t$ is the hopping 
integral) may provide the possibility for interatomic distance manipulation 
due to the molecule instability. \vspace{-2em}
\end{abstract}
\vspace{-2em}


\end{frontmatter}

Recent progress in nanotechnology made possible to fabricate small systems 
attached to leads, and to measure the conductance of even a single hydrogen 
molecule \cite{smit}. 
In such systems strong electron correlations can play 
a decisive role \cite{liang}.
The theoretical understanding of the experiments is far from complete,
partly because the nanoscale leads geometry often introduces a degree of 
irreproducibility, and partly due to many body effects on the intrinsic 
transport properties of such single--molecule devices. In particular,
the electron--phonon coupling was recently shown shrink the conductance
peak for a single quantum dot in a Kondo regime \cite{corn}, since the
on--site Coulomb repulsion $U$ is effectively renormalized in a~system
ground state. Here we describe a qualitatively similar effect in a double 
quantum dot, produced solely by electron--electron interaction only through 
a~completely different physical mechanism: the correlation--induced 
wave--function renormalization \cite{spary}.

\begin{figure}[!b]
\begin{center}
\includegraphics[width=\columnwidth]{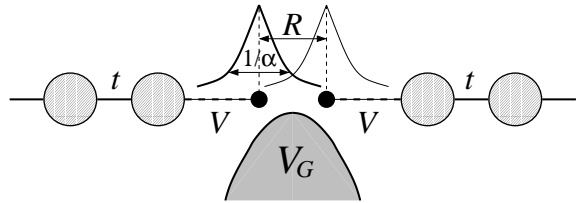}
\end{center}
\caption{
  Diatomic molecule as a double quantum dot.}
\label{qd2fig}
\end{figure}

We start with the Hamiltonian which is a generalization of the 
Anderson impurity model \cite{ander}, and can be written as
\begin{equation}
\label{ham5}
  H = H_L+V_L+H_C+V_R+H_R,
\end{equation}
where $H_C$ models the central region, 
$H_{L(R)}$ describes the left (right) lead, and $V_{L(R)}$ is the coupling
between the lead and the central region.
Both $H_{L(R)}$ and $V_{L(R)}$ terms have a tight--binding form, with the 
hopping $t$ and the tunneling amplitude $V$, as depicted schematically 
in~Fig.\ \ref{qd2fig}.
The central--region Hamiltonian $H_C$ describes a double quantum dot with 
electron--electron interaction, and has a form
$$
  H_C = (\epsilon_a-eV_G)\sum_{j=1,2}n_i
  -t'\sum_{\sigma=\uparrow,\downarrow}(c_{1\sigma}^{\dagger}c_{2\sigma}
 +\mbox{h.c.})
$$
\vspace{-1.5em}
\begin{equation}
\label{hamc}
 + U\sum_{j=1,2}n_{i\uparrow}n_{i\downarrow} + Kn_1n_2 + (Ze)^2/R,
\end{equation}
where $\epsilon_a$ is atomic energy, $V_G$ is an external gate voltage,
$t'$ is the internal hopping integral, $U$ and $K$ represents the
intra-- and inter--site Coulomb interactions, respectively, and the last term 
describes the Coulomb repulsion of the two ions at the distance $R$. 
Here we put $Z=1$ and calculate all the parameters $\epsilon_a$, $t'$, $U$, 
and $K$ as the Slater integrals \cite{slat} for $1s$--like hydrogenic orbitals
$\Psi_{1s}(\mathbf{r})=\sqrt{\alpha^3/\pi}\exp(-\alpha|\mathbf{r}|)$,
where $\alpha^{-1}$ is the orbital size (\emph{cf.}\ Fig.\ \ref{qd2fig}).
The parameter $\alpha$ is optimized to get a minimal ground state energy
for whole the system described by the Hamiltonian (\ref{ham5}). Thus,
following the idea of EDABI method \cite{ryspa}, we reduce the number
of physical parameters of the problem to just a three: the interatomic 
distance $R$, the gate voltage $V_G$, and the lead--molecule coupling
$V$ (we put the lead hopping $t=1\ \mathrm{Ry} = 13.6\ \mathrm{eV}$ to work in 
the wide--bandwidth limit).

\begin{figure}[!t]
\begin{center}
\includegraphics[width=\columnwidth]{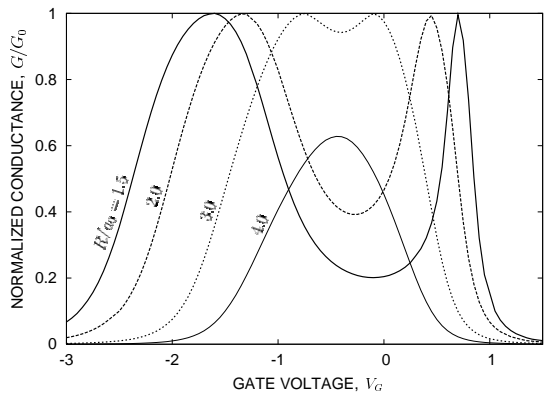}
\end{center}
\caption{
  Zero--temperature conductance of the system in Fig.\ \ref{qd2fig}
  as a function of the gate voltage $V_G$ and interatomic distance $R$.
  $1s$ orbital size $\alpha^{-1}$ is optimized variationally. 
  The lead paremeters are $t=1\mathrm{ Ry}$ and $V=0.5t$.
}
\label{gH2fig}
\end{figure}

We discuss now the molecule conductivity calculated from the Rejec--Ram\v{s}ak 
two--point formula \cite{reram}:
\begin{equation}
  G=G_0\sin^2\pi[E(\pi)-E(0)]/2\Delta,
\end{equation}
where $G_0=2e^2/\hbar$ is the conductance quantum,
$\Delta=1/N\rho(\epsilon_F)$ is the average level spacing at Fermi energy,
determined by the density of states in an infinite lead $\rho(\epsilon_F)$,
$E(\pi)$ and $E(0)$ are the ground--state energies of the system with periodic
and antiperiodic boundary conditions, respectively. 
$E(\phi)$ is calculated for $\phi=0,\pi$ within the Rejec--Ram\v{s}ak 
variational method \cite{reram}, complemented by the inverse orbital size 
$\alpha$ optimization, as mentioned above.
We use typically $N=10^2\div 10^3$ sites to reach the convergence.

In Fig.\ \ref{gH2fig} we show the conductivity for $V=0.5t$, and different
values of the interatomic distance $R$. The conductance spectrum evolves from 
the situation well--separated peaks corresponding to the independent filling
of bonding and antibonding molecular orbitals ($R\lesssim 2a_0$, where 
$a_0$ is the Bohr radius), to the single peak in the intermediate range 
($R\approx 3a_0$), which decays when $t'\ll V$ for large $R$.

\begin{figure}[!t]
\setlength{\unitlength}{0.01\columnwidth}
\begin{picture}(100,50)
\put(-8,0){\includegraphics[width=1.2\columnwidth]{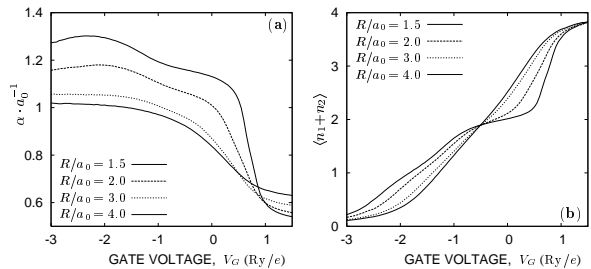}}
\end{picture}
\caption{
  The optimized inverse inverse odbital size $(a)$ and average central
  region occupancy $(b)$ for the diatomic molecule attached to noninteracting
  leads characterized by the parameters $t=1\mathrm{ Ry}$ and $V=0.5t$.
}
\label{AlpNjfig}
\end{figure}

Probably, the most interesting feature of the spectra depicted in 
Fig.\ \ref{gH2fig} is their strong asymmetry for small $R$. 
Namely, the low--$V_G$ conductance peak,
corresponding to the system filling $\langle n_1\!+\!n_2\rangle\approx 1$ 
(one \emph{hole}) is significantly wider than the high--$V_G$ peak for
$\langle n_1\!+\!n_2\rangle\approx 3$ (one \emph{extra electron}).
Such a particle--hole symmetry breaking is the novel feature, observed
when including the correlation--induced basis optimization, and absent
in the parametrized--model approach \cite{reram,bulk}. 
It is also a new feature of 
a nanosystem, do not observed in mesoscopic double quantum dots \cite{kouw},
where the particle--hole symmetry is perfect.

The relation between the observed asymmetry, basis renormalization, 
and electron correlations can be clarified as follows.
\emph{First}, one can observe that the optimal values of the variational 
parameter $\alpha$, provided in Fig.\ \ref{AlpNjfig}a, decrease dramatically
for high $V_G$, corresponding to the overdoped situation 
$\langle n_1\!+\!n_2\rangle >2$ (\emph{cf.}\ Fig.\ \ref{AlpNjfig}b).
This is because the system minimize energy of double occupancies, which
is of the order $U\sim\alpha$ \cite{slat}.
\emph{Then}, we focus on an small $R$ limit, in which the well separation
of molecular orbitals allows one to approximate the expression for a single
impurity at zero temperature  \cite{meir}
\begin{equation}
\label{gsingle}
  G=G_0\sin^2(\pi\langle n_k\rangle/2),
\end{equation}
where $\langle n_k\rangle$ is the bonding ($k=0$) or antibonding ($k=\pi$)
orbital occupancy. Expanding around the maximum we have
$G(V_G)-G(V_G^{*,k})\approx -G_0(\chi_c\pi/2)^2(V_G-V_G^{*,k})$, where
$V_G^{*,k}$ ($k=0,\pi$) is the low/high voltage peak position and $\chi_c$
is the charge susceptibility, which may be approximated as 
$\chi_c\approx\partial\langle n_1\!+\!n_2\rangle/\partial V_G$, since
the orbitals are filled separately. Values of the later derivative read
from Fig.\ \ref{AlpNjfig}b around the low and high voltage peak positions 
($\langle n_1\!+\!n_2\rangle\approx1$ and $\approx 3$, respectively) provides
a clear asymmetry. Moreover, the expansion of $G(V_G)$ allows one to roughly
estimate the peak width as $\Delta V_G\approx\chi_c^{-1}\approx (U+K)/2
\sim\alpha$, what provides an indirect correspondence between the spectrum 
asymmetry and basis renormalization.

\begin{table}[!b]
\caption{
  The binding energy $\Delta E$ and the bond length $R_{\min}$ for different 
  coupling to the lead $V$. The conductance at the energy minimum is also 
  provided.\label{stabtab}
}
\begin{center}
\begin{tabular}{rrrr}
\hline\hline
$V/t$ & $\mbox{ }\Delta E\mbox{\hspace{2ex}}$ & $\mbox{ }R_{\min}/a_0$
  & $\mbox{ }G_{\min}^{V_G=0}/G_0$ \\ 
\hline
0.0  & -0.296  & 1.43  & 0 \vspace{-1em}\\
0.1  & -0.293  & 1.44  & 10$^{-4}$ \vspace{-1em}\\
0.2  & -0.282  & 1.46  & 0.004 \vspace{-1em}\\
0.3  & -0.180  & 1.52  & 0.023 \vspace{-1em}\\
0.4  & -0.040  & 1.59  & 0.093 \vspace{-1em}\\
0.5  & +0.074  & 1.93  & 0.422 \vspace{0em}\\ 
\hline\hline
\end{tabular}
\end{center}
\end{table}

The practical possibility of the conductance measurement involving atoms
manipulation (changing of $R$) were also explored in terms of the system
stability. Namely, the binding energies 
$\Delta E\equiv E_{R_{\min}}-E_{\infty}$ (where $R_{\min}$ is the bond length),
listed in Table \ref{stabtab}, shows that the system become metastable around
$V=0.5t$ ($\Delta E>0$ indicates a \emph{local} energy minimum).
Therefore, the binding of the atoms to the lead become stronger than 
interaction between the atoms, that may allow individual atoms manipulation.

In \emph{summary}, we have supplemented the double quantum dot conductance 
calculation with the single--particle basis optimization. 
Apart from simplifying the analysis, such an approach leads to 
substantially new physical effects, since it breaks the particle--hole
symmetry of the Hamiltonian.

The work was supported by the Polish Science Foundation (FNP)
and by Polish Ministry of Education and Science, Grant No.\ 1 P03B 001 29.

\end{document}